\documentstyle[11pt,a4,epsfig]{article}
\begin{document}
\baselineskip 12pt
%\begin{flushright} {\bf OCHA-PP 129} \end{flushright}
\begin{center}{\Large \bf
On Spectrum of Extremely High Energy Cosmic Rays through \\
Decay of Superheavy Particles}\\

\medskip
{Y\={u}ichi Chikashige$^1$  and Jun-ichi Kamoshita$^2$ }

\smallskip
{ $^1$\it Faculty of Engineering, Seikei University,
              Musashino, Tokyo 180-8633, Japan \\
  $^2$\it Department of Physics, Ochanomizu University,
              2-1-1 Otsuka, Bunkyo, Tokyo 112-8610, Japan}
\end{center}

%\title{On Spectrum of Extremely High Energy Cosmic Rays through \\
%Decay of Superheavy Particles}
%\author{Y\={u}ichi Chikashige$^1$  and Jun-ichi Kamoshita$^2$ }
%%\address{
%\date{
%   $^1$\it Faculty of Engineering, Seikei University,
%              Musashino, Tokyo 180-8633, Japan \\
%   $^2$\it Department of Physics, Ochanomizu University,
%              2-1-1 Otsuka, Bunkyo, Tokyo 112-0012, Japan}  
%\date{\today}
%
%\maketitle

\smallskip

\begin{abstract} 
 We propose a formula for flux of extremely high energy cosmic rays (EHECR)  
 through decay of superheavy particles.
 It is shown that EHECR spectrum reported by AGASA is reproduced by the formula.
 The presence of EHECR suggests, according to this approach,
 the existence of superheavy particles
 with mass of about $7 \times 10^{11}$GeV
 and the lifetime of about $10^9$ years.
 Possibility to obtain a knowledge of $\Omega_0$ of the universe
 from the spectrum of EHECR is also pointed out.
\end{abstract}

%\vskip 0.4truecm
%\pacs{
%}
% e-mail: $^1$ chika@ge.seikei.ac.jp    $^2$ kamosita@theory.kek.jp 

\vskip 0.5truecm
%\twocolumn
 
  The unexpected energy spectrum of
 the extremely high energy cosmic rays (EHECR)
 with the energy above $10^{19.8}$eV has been reported by 
 Akeno Giant Air Shower Array (AGASA) collaboration
 which used the updated data set\cite{AGASA98}.
 The existence of EHECR has been known for about 30 years \cite{Hillas},
 among which the highest energy of EHECR,
 $(3.2\pm0.9)\times 10^{20}$eV, was recorded in Fly's Eye \cite{FE}.
 When we regard nucleons or nuclei as constituents of EHECR,
 the attenuation length is estimated less than $\sim 100$ Mpc
 due to Greisen-Zatsepin-Kuz'min (GZK) effect \cite{{GZK},{gzk-2-5}}.
 Thus if distances from Earth to any sources of EHECR are over 100 Mpc,
 it is difficult to explain EHECR in terms of nucleons or nuclei.
 One may expect
 that sources of EHECR would exist within about 100 Mpc from Earth, 
 but such sources have not been found.
 Here we can make a brief list of yet unsolved problems on EHECR
 as follows: 
 (1) the chemical compositions of EHECR are unknown, 
 (2) the sources of EHECR are  unclarified, and
 (3) the shape of energy spectrum reported by AGASA is unexplained.

   In order to solve these problems,
 several ideas have been proposed.
 There exist mainly two approaches based
 on astrophysical aspect and particle physical aspect.
 From the astrophysical aspect,
 number of acceleration mechanisms of the ordinary particles
 have been proposed.
 For example, the protons are accelerated to the extremely high energy
 by the relativistic jets from AGN \cite{mannheim}.
 The production of EHECR by the Gamma-Ray Bursts is
 also considered \cite{BW}. 
 There are, however,
 no particular astronomical objects
% pointed
 in the directions of EHECR \cite{{AGASA98},{FE}}.
 
 On the other hand, from the particle physical aspect,
 the production mechanism of EHECR
 has been proposed \cite{r-3}.
 A basic idea is that EHECR are produced
 by decay of superheavy particles. 
 In this case, EHECR are consisted of
 either the standard particles (proton, photon or neutrino)
 or the new particles (for example, neutralino or gluino).
 Furthermore, instead of decay products of superheavy particles,
 quasi-stable superheavy particles can also be considered
 as constituent of EHECR in itself. 
 For example, colored monopoles
 have been examined in ref.\cite{p-15}.

 When decaying superheavy particles are considered as sources of EHECR,
 it is plausible to imagine
 the mass of the superheavy particles larger than about $10^{12}$GeV, 
 since the highest energy of EHECR is $(3.2\pm0.9)\times 10^{20}$eV.
 There are many models in particle physics
 which introduce such superheavy particles,
 for example,
    supersymmetric grand unified theory
 or the see-saw model for neutrino mass.
 However, the superheavy particles that should decay into EHECR
 have to have a long lifetime
 which is nearly or greater than the present age of the universe,
 because, if the superheavy particles decay fast, 
 the remnants of the superheavy particles are very little,
 and as a result, we cannot observe EHECR produced via the decay.
 Although the order of lifetime is usually proportional
 to the inverse of mass of the particle,
 some kind of models \cite{NY} permits quasi-stable superheavy particles
 contrary to their heaviness.

 It is attractive to make clear of the relation
 between the EHECR problem and the particle physical aspect,
 because we can expect
 that the feature of EHECR leads us to ones
 beyond the standard model of particle physics
 with a large mass scale larger than about $10^{12}$GeV. 

 In this paper,
 we examine the problem of EHECR from the particle physical aspect, 
 although we do not reject at the present time
 a possibility to explain EHECR from
 astronomical origins like quasisteller objects
 which are recently considered seriously in ref.~\cite{FB}.
 We propose a proper formula for the flux of EHECR in the beginning. 
 This formula should be used
 when EHECR are produced via the decay of the superheavy particles.
 Reproducing the spectrum of EHECR observed at AGASA by using our formula,
 we show that the lifetime of the superheavy particles is suggested
 to be about 10\% of the Hubble time
 when their mass is $7\times10^{11}$GeV.
 Furthermore, it is also pointed out
 that the possibility to obtain a knowledge of the omega parameter,
 $\Omega_0$, of the universe from the energy spectrum of EHECR. 

 At first, we show
 the formula for the flux to the cosmic rays
 as decay products of superheavy particle $X$.
 Let $f$ be the particle in decay products of $X$ and
    produce Extensive Air Showers (EAS) when it reaches to the atmosphere.
 We denote the number density of the source at time $t_e$ as $n_X(t_e)$
 and the partial decay constant for a decay mode including $f$ as $\Gamma_f$.
 Then the production rate of $f$ at time $t_e$ is
 determined by $\Gamma_f n_X(t_e)$.
 The general expression of the flux of cosmic rays,
 after taking average about the angular distribution,
 is given in the Robertson-Walker metric with the scale factor $a(t)$
 as follows:
 \begin{eqnarray}
   J(E_{obs}) &=&
        \frac{1}{4\pi}\int^{t_0}_{t_{min}}{\rm d}t_e\ 
                 \left(\frac{a(t_e)}{a(t_0)}\right)^3
                 n_X(t_e)   \frac{{\rm d}\Gamma_f}{{\rm d}E_e}
                 \frac{{\rm d}E_e}{{\rm d}E_{obs}}~,
%\nonumber \\  & & 
 \label{eqn:genflx}
\end{eqnarray}
 where 
 $E_{obs}$ is the observed primary energy of cosmic rays,
 and
 ${\rm d}\Gamma_{f}/{{\rm d}E_e}$ is the energy distribution of cosmic rays
 emitted by the source at time $t_e$.
 The present time is represented by $t_0$,
 while $t_{min}$, is determined by a physical condition
 that $t_0 - t_{min}$ is smaller than an attenuation time of $f$.

The number density of the superheavy particles $n_{X}(t)$ is evolved
 by the following Boltzman equation,
\begin{eqnarray}
    \frac{{\rm d}(a(t)^3n_{X}(t))}{{\rm d}t}= - \Gamma_X  a(t)^3 n_{X}(t),
 \label{eqn:Bltz}
\end{eqnarray}
 where $\Gamma_X$ is total decay width of $X$.
 This equation gives
\begin{eqnarray}
   n_X(t) = \left(\frac{a(t_0)}{a(t)}\right)^3
              n_X(t_0) \exp[\Gamma_X(t_0-t)]\,.
 \label{eqn:nxt}
\end{eqnarray}
 By substituting eq.~(\ref{eqn:nxt}) into eq.~(\ref{eqn:genflx}), 
 we obtain 
\begin{eqnarray}
   J(E_{obs}) &=& \frac{n_X(t_0)}{4\pi}\int^{t_0}_{t_{min}}{\rm d}t_e
                   \exp[\Gamma_X(t_0-t_e)] 
                    \frac{{\rm d}\Gamma_f}{{\rm d}E_e}
                    \frac{{\rm d}E_e}{{\rm d}E_{obs}}~. 
% \nonumber \\
    \label{eqn:flx}
\end{eqnarray}
 Here we would like to make a comment on this formula.
 In the previous works,
 similar formulae have been used \cite{{r-3},{YDJS}};
 however, 
 the effect of the decrease
 in the number density of parent particles has been neglected,
 since their lifetime has been assumed
 to be greater than the age of the universe.
  On the other hand, in eq.~(\ref{eqn:flx}), 
 this effect is included in the form of exponential damping factor,
 $ \exp[\Gamma_X(t_0-t)] $,
 since $X$ is supposed to have the same order lifetime
 as the present age of the universe.
 We should use eq.~(\ref{eqn:flx}) rather
 than the formula used in previous works
 when we estimate the flux of EHECR produced
 by the decay of superheavy particles.
  This is because the number density of parent particles decreases
 due to their decay.
 This effect is very important to reproduce the spectrum of EHECR
 as shown later in this paper.

  Our formula, eq.~(\ref{eqn:flx}),
 can be applied to the case
 that $f$ can travel almost freely through the cosmic background radiation,
 though the attenuation due to the GZK effect has been neglected.
 For example, we can take $f$ to be neutrino or neutralino.
 The difference between them is
 the translation rate $R_{c2a}$ of EHECR to EAS
 when they reach at Earth and produce EAS.
\begin{eqnarray}
  R_{c2a}&\equiv&\frac{\rm number\ of\ EHE\ air\ shower \ events}
                      {\rm number\ of\ EHECR\ incident\ on\ Earth}~.
\end{eqnarray}
 Then the flux of EAS, $J(E_{obs})|_{\rm EAS}$,
 is derived from the flux of EHECR,  
 $J(E_{obs})|_{\rm EHECR}$, as
\begin{eqnarray}
      J(E_{obs})|_{\rm EAS} = J(E_{obs})|_{\rm EHECR} R_{c2a}.
 \label{eqn:rc2a} 
\end{eqnarray}
 Neutrino gives the translation rate larger value as an order of
 $ R_{c2a} =10^{-6} $ \cite{ghandi} than neutralino.

 Now we consider a case
 that EHECR are taken to be neutrinos
 which are produced by the two-body decay of $X$.
 Its mass, $M_X$, can be expected about twice the highest energy of EHECR.
 Since the highest energy recorded by Fly's Eye is
                     $(3.2\pm 0.9)\times10^{20}$eV,
 we take $M_X$ as $7\times 10^{11}$GeV in the following.
 The energy spectrum right at the emitted point is
 to be the monochromatic one in the two-body decay,
\begin{eqnarray}
    \frac{{\rm d}\Gamma_f}{{\rm d}E_e}=\Gamma_f\,\delta(E_e-M_X/2).
\label{eqn:monodep}
\end{eqnarray}
 
 Hereafter, for simplicity,
 we limit ourselves to consider
 such a case that there is unique channel for
 two-body decay, {\it i.e.} $ X \rightarrow \nu \,+$ some particle,
 and then $\Gamma_f$ becomes
 just the total decay width $\Gamma_{tot}$.
 Since the kinematics tells
 that the energy spectrum is monochromatic in the present case, 
 we can obtain the energy distribution of EHECR
 without the detailed description of the interaction.
 Thus we will be able to discuss the problem of EHECR
 rather in the model-independent way.

 The red-shift relation between $E_e$ and $E_{obs}$
 is represented as $(1+z)E_{obs}=E_e$.
 Then the monochromatic energy condition becomes
\begin{eqnarray}
    \delta(E_e-M_X/2)=\frac{1}{E_{obs}}
                      \frac{1}{|{\rm d}f(t)/{\rm d}t|}
                      \delta(t-t_{\alpha})\theta(z),
  \label{eqn:dltime}
\end{eqnarray}
 where the step function $\theta(z)$ is necessary
 to satisfy the condition $z \geq 0$.
 In eq.~(\ref{eqn:dltime}), $f(t)\equiv 1+z(t)$, 
 and $t_{\alpha}$ is defined as a solution of 
 $f(t_{\alpha})={M_X}/{2E_{obs}}$.
 The expression of $f(t)$ is different for each of three cases,
 open universe of $\Omega_0<1$,
 flat universe of $\Omega_0=1$ and
 closed universe of $\Omega_0>1$. 
 By using eqs.~(\ref{eqn:flx})~$\sim$~(\ref{eqn:dltime}),
 we can calculate the energy spectrum of EHECR.
 For the moment, we consider the case for $\Omega_0=1$,
 where $f(t)=(t_0/t)^{2/3}$.
 In this case, the spectrum of EAS is given by
\begin{eqnarray}
  J(E_{obs})\mid _{\rm EAS} &=&
             R_{c2a} n_X(t_0) \frac{ \sqrt{2} \Gamma_X }{2 \pi H_0}
                              \frac{\sqrt{E_{obs}}}{{M_X}^{3/2}}
%\times \nonumber \\  & & 
             \exp\left\{ \frac{2 \Gamma_X}{3 H_0}
                         \left[ 1-\left(\frac{2 E_{obs}}{M_X}\right)^{3/2} \right]
                \right\}~.
  \label{eqn:lgflx}
\end{eqnarray}
 The AGASA's data above the GZK cutoff \cite{AGASA98} distributes
 over the energy region from $10^{19.8}$eV to $10^{20.5}$eV,
 which corresponds to the change of 1 to 5 for $1 + z$.
 This means that
 the most remote position of $X$ is about 7 Gpc in flat universe case.
 This distance is certainly shorter
 than the mean free length of supposed neutrinos
 on the collisions with background photons or neutrinos \cite{tmin}.
 Thus we can practically put $t_{min}$ into 0 in eq.~(\ref{eqn:flx}).  
 We see from eq.~(\ref{eqn:lgflx})
 that the spectrum of EHECR is determined by
 the three parameters $M_X$, $r\equiv\Gamma_X/H_0$, and $R_{c2a}n_X(t_0)$.
 Since $R_{c2a}$ appears as a product with $n_X(t_0)$ in eq.~(\ref{eqn:lgflx}),
 fortunately an ambiguity of $R_{c2a}$ does not affect
 our estimate of the spectrum.
  When we reproduce the spectrum reported by AGASA, 
 we must multiply the factor $E_{obs}^3$ to eq.~(\ref{eqn:lgflx}).

\begin{figure}[h]
   \epsfxsize = 14 cm
   \centerline{\epsfbox{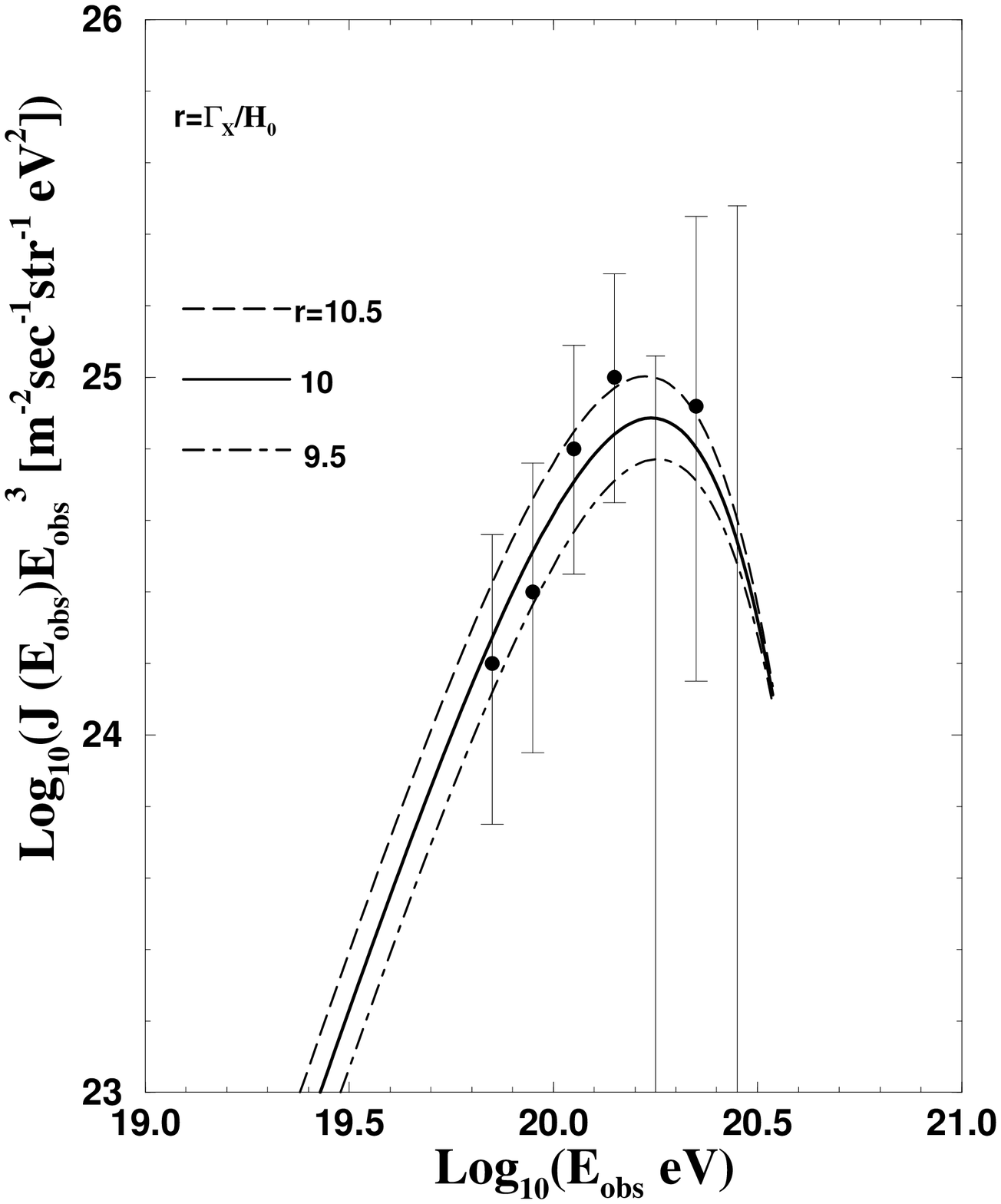}}
 \caption[]{
          Curves of $J(E_{obs})E_{obs}^3$ 
        for the two-body decay case.
        Data points and error bars are taken from AGASA data given in \cite{AGASA98}.
        Input parameters are taken as follows;
        $M_X = 7\times 10^{11}$GeV,
        $r\equiv \Gamma_X/H_0 = 9.5, 10$, and 10.5,
        $\Omega_0 = 1$,
        $\rho_X(t_0) = M_X n_X(t_0) = 10^{-8} \rho_{cr} (t_0)$, and 
        $R_{c2a} = 10^{-6}$. 
           }
 \label{fig:2bd}
 \end{figure}

 Fig.~\ref{fig:2bd} displays the curves of $\log_{10}(J(E_{obs})E_{obs}^3)$
 for the case that EHECR are to be neutrinos produced by the two-body decay of $X$.
 The flux $J(E_{obs})$ is calculated by eq.~(\ref{eqn:lgflx}).
 The end point of the energy spectrum locates at $M_X/2$.
 The energy of EHECR produced at time $t_{\alpha}$ is lowered
 to $E_e/(1+z)$ by the red shift.
 The peak around $10^{20}$eV is reproduced
 by synergistic effect between the decreasing factor of $X$,
 $\exp [\Gamma_X (t_0 - t_1 )]$,
 and the expanding factor of the universe, 
 $({t_0}/{t_1}) ^{2/3}$ in eq.~(\ref{eqn:flx}). 
 We see remarkable conformity of the energy spectrum of EHECR
 by eq.~(\ref{eqn:lgflx}) with the one reported by AGASA.

 The location of peak of the spectrum shown in Fig.~\ref{fig:2bd}
 is derived from eq.~(\ref{eqn:lgflx}) as
 \begin{eqnarray}
    E_{obs}({\rm peak})=\frac{M_X}{2}\left(\frac{3.5}{r}\right)^{2/3}.
 \end{eqnarray}
 In order that the peak appears in Fig.~\ref{fig:2bd}, $r>3.5$ is needed.
 We see from the data point given by AGASA
 that the peak seems to locate at $E_{obs}\sim 10^{20}$eV,
 though this location is vague in the data.
 When the inequalities $20.1<\log_{10}E_{obs}<20.4$ are imposed,
 we obtain
 \begin{eqnarray}
    3.5\left(\frac{M_X}{2}10^{-11.4}\right)^{1.5}< r <
    3.5\left(\frac{M_X}{2}10^{-11.1}\right)^{1.5}
 \end{eqnarray}
 where $M_X$ is given in the unit of GeV.
 For $M_X=7\times10^{11}$GeV, we find $5.8 < r < 16$.
 The inequalities turn the lifetime of $X$ into
 $\tau_X= (0.09\sim0.3)\,t_0$.
 
 When we see AGASA's data \cite{AGASA98} in detail,
 there is no event in the energy bin between $10^{20.2}$ and $10^{20.3}$ eV.
 We anticipate, however, appearing of one event at this energy bin
 from our present consideration. 
 But if there remains this vacancy in future,
 this spectrum should be interpreted from our standpoint
 as a suggestion
 that two kinds of superheavy particles with different masses exist.
 The spectrum below $10^{20.2}$ eV is
 due to superheavy particles with mass of $3 \times 10^{11}$GeV,
 while the one over $10^{20.3}$eV is
 due to another superheavy particles with mass of $7 \times 10^{11}$GeV.

\begin{figure}[h]
    \epsfxsize = 14 cm
    \centerline{\epsfbox{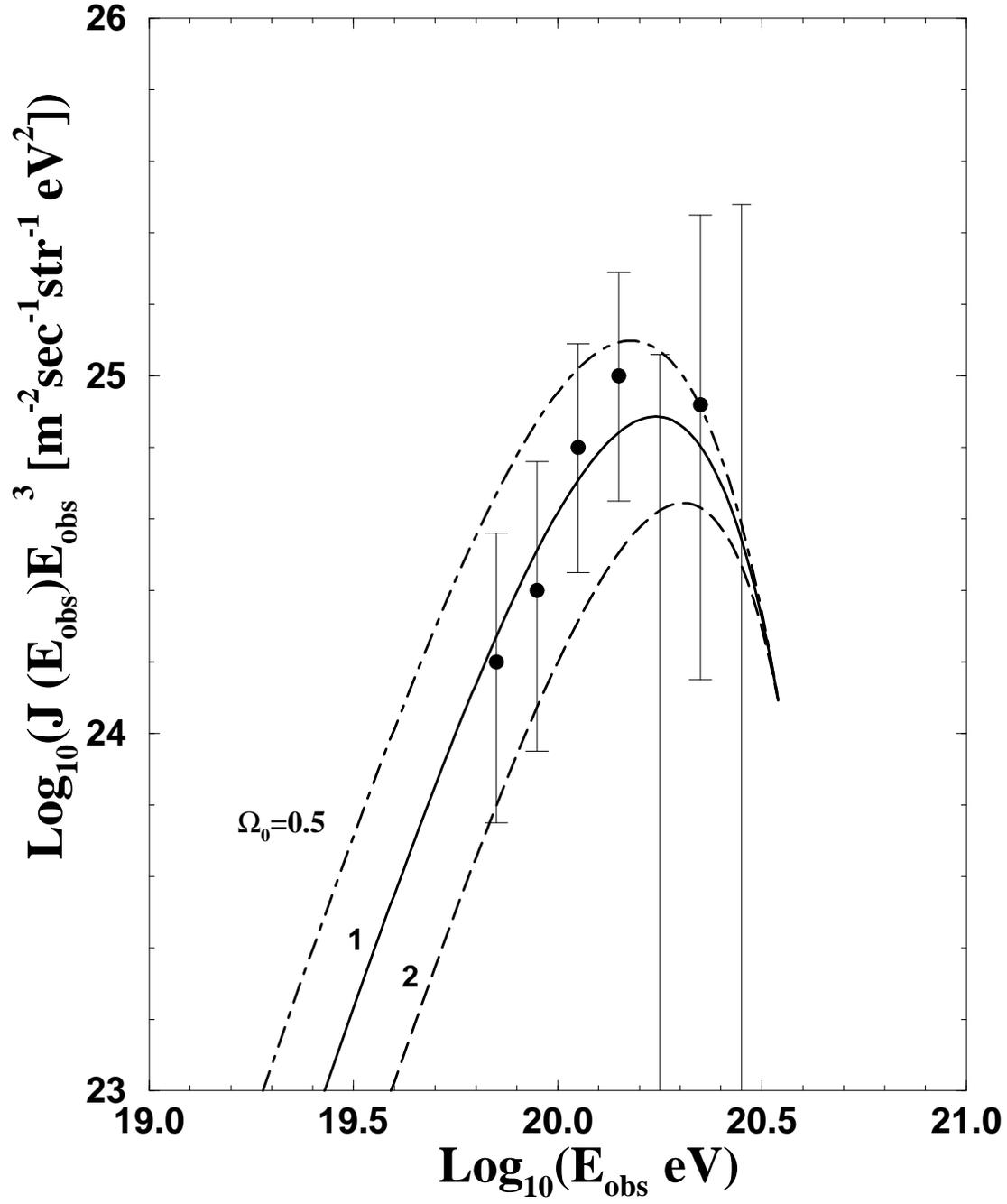}}
 \caption[]{ Curves of $J(E_{obs})E_{obs}^3$ 
        for the two-body decay case.
        We take the omega parameters
        as $\Omega_0=0.5, 1, 2$, and $r = 10$.
        Other input parameters are taken
        to be the same as in Fig.~\ref{fig:2bd}.}
 \label{fig:omega}
 \end{figure}

 In Fig.~\ref{fig:omega},
 the curves of $\log_{10}(J(E_{obs})E_{obs}^3)$ are shown
 for $\Omega_0=0.5, 1$, and 2.
 We put $r = 10$ for $M_X = 7\times10^{11}$GeV as an example.
 Statistics of the current data above $10^{19.8}$ eV is not enough
 to derive information of $\Omega_0$.
 We hope from this point
 that the statistics of the EHECR observation increases in future.

\begin{figure}[h]
    \epsfxsize = 14 cm
    \centerline{\epsfbox{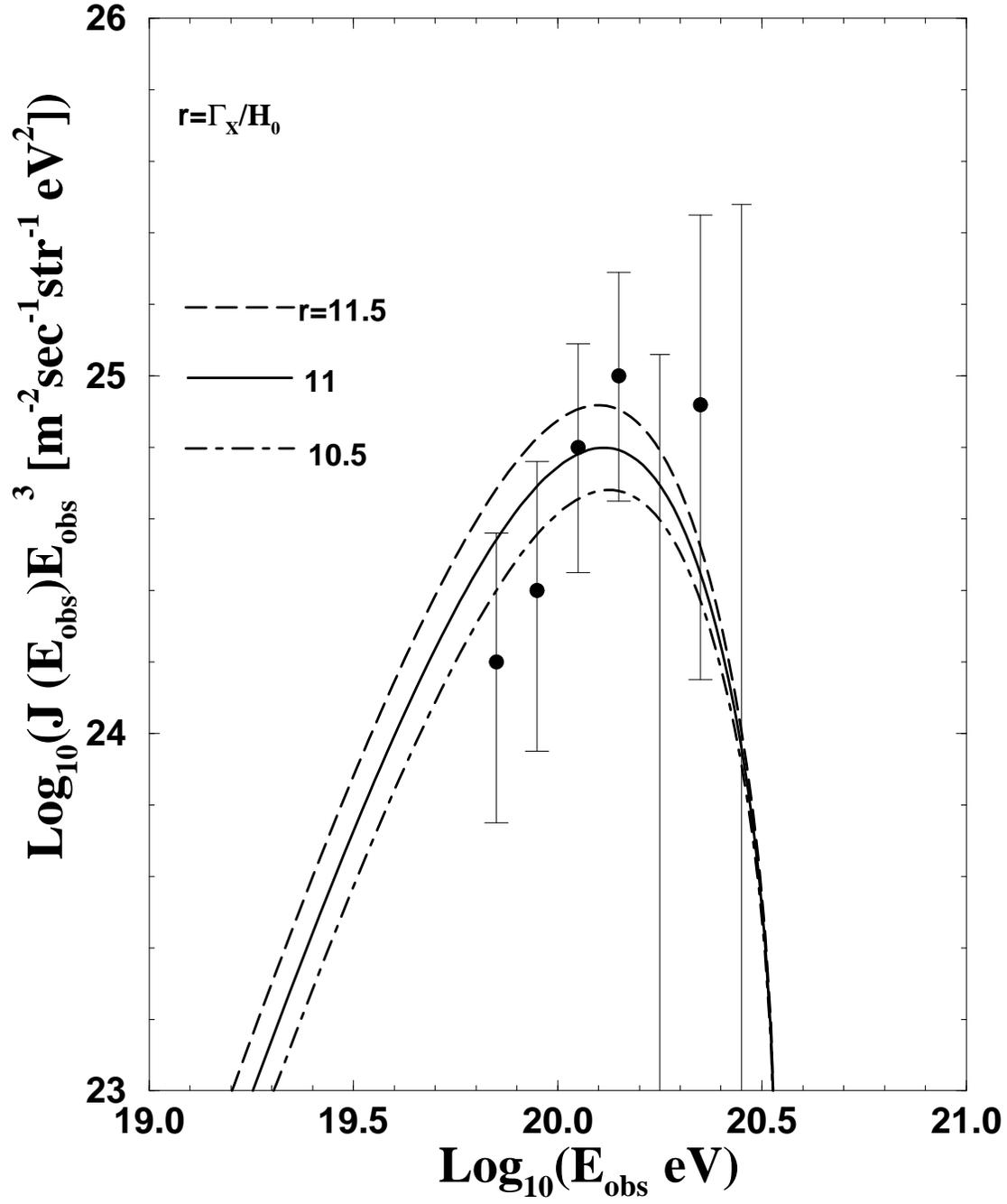}}
 \caption[]{ Curves of $J(E_{obs})E_{obs}^3$ 
        for the three-body decay case.
        We take the ratio of $\Gamma_X/H_0$ as $r = 10, 10.5$, and 11.5.
        Other input parameters are taken
        to be the same as in Fig.~\ref{fig:2bd}.}
 \label{fig:3bd}
\end{figure}

 So far we have considered two-body decay case.
 Here we address a question
 how the energy spectrum changes for multi-body decay case.
 As an example,
 we show the curves of
 $\log_{10}(J(E_{obs})E_{obs}^3)$ for three-body decay
 due to four-Fermi interaction.
 We can see from Fig.~\ref{fig:3bd}
 that the curves for three-body decay case
 do not reproduce the data so well
 as those for two-body decay case in Fig.~\ref{fig:2bd}.

 In summary,
 we have examined the problem of EHECR from the particle physical aspect,
 proposing a suitable formula to the flux of the cosmic rays
 through decay of superheavy particles. 
 Then the energy spectrum of EHECR has been reproduced satisfactorily
 as that of the neutrinos via two-body decay of $X$
 with mass of $7 \times 10^{11}$GeV and lifetime of around $0.1\,t_0$.
 $R_{c2a} \Omega_X (t_0) \sim 10^{-14}$ has been required
 to reproduce the energy spectrum of EHECR,
 where $\Omega_X \equiv \rho_X /\rho_{cr}$ as is usually defined.
 Our analysis is not confined exclusively to the neutrino case as mentioned before.
 We can take neutralino instead of neutrino.
 The difference between neutralino and neutrino is only the value of $R_{c2a}$,
 and $R_{2ca}$ becomes smaller for neutralino.
 Thus, $\Omega_X (t_0)$ turns out to be much larger
 for neutralino case than $10^{-8}$ for neutrino case.
 Furthermore,
 we have the possibility to derive a knowledge of the omega parameter,
 $\Omega_0$, of the universe from the energy spectrum of EHECR.
 We expect to acquire further information of superheavy particles as well
 as $\Omega_0$ of the universe from future observations
 where more data will be accumulated to give higher statistics.
 From this aspect, future detectors like the HiRes, the Telescope
 Array, the Pierre Auger, and the Owl \cite{Physics Today} are greatly awaited.

  One of the authors
 (Y. C.) thanks Tetsuya Hara for his kind correspondence.

%%%%%%%%%%%%%%%%%%%%%%%%%%%%%%%%%%%%%%%%%%
%%%%%%
%%%%%%       References
%%%%%%
%%%%%%%%%%%%%%%%%%%%%%%%%%%%%%%%%%%%%%%%%%
%\vspace{20pt}

%

%%%%%%%%%%%%%%%%%%%%%%%%%%%%%%%%%%%%%%%%%%
%%%%%%
%%%%%%       Figures
%%%%%%
%%%%%%%%%%%%%%%%%%%%%%%%%%%%%%%%%%%%%%%%%%
\newpage


\end{document}